\documentclass[superscriptaddress,
amsmath,amssymb,
aps,
pra,
floatfix,
twocolumn]{revtex4-2}

\usepackage[utf8]{inputenc}
\usepackage[english]{babel}
\usepackage[T1]{fontenc}
\usepackage{hyperref}
\usepackage{amsmath}
\usepackage{amssymb}
\hypersetup{colorlinks=true,
    linkcolor=blue,
    filecolor=blue, 
    citecolor=blue,
    urlcolor=blue,
}

\usepackage{braket, hyperref, mathtools, verbatim}
\usepackage{physics}
\usepackage{graphicx}
\usepackage{braket}
\usepackage{dcolumn}
\usepackage{bm}
\usepackage{tabularx}
\usepackage{makecell}
\usepackage{tikz}
\usepackage{svg}
\usepackage{lipsum}
\usepackage{bbold}
\usepackage{algorithm2e}
\usepackage{enumerate}
\usepackage{mathtools}
\SetKwComment{Comment}{/* }{ */}
\RestyleAlgo{ruled}
\SetAlgoCaptionLayout{raggedright}

\newcommand{\sH}{\mathcal{H}}
\newcommand{\sO}{\mathcal{O}}
\newcommand{\sA}{\mathcal{A}}

\newcommand{\had}[1][ ]{\hat{a}^{\dagger #1}}
\newcommand{\ha}[1][ ]{\hat{a}^{#1}}
\newcommand{\hbd}[1][ ]{\hat{b}^{\dagger #1}}
\newcommand{\hb}[1][ ]{\hat{b}^{#1}}

\newcommand{\hSig}[0]{\hat{\sigma}} 

\newcommand{\hHp}{\hat{H}_\text{PA}}

\newcommand{\hU}{\hat{U}}
\newcommand{\hH}{\hat{H}}
\newcommand{\hO}{\hat{O}}
\newcommand{\hs}{\hSig}
\newcommand{\hz}{\hat{z}}
\newcommand{\hx}{\hat{x}}
\newcommand{\hy}{\hat{y}}
\newcommand{\HC}{\textit{h.c.}}

\begin{document}

\title{Parametric Amplification of Spin-Motion Coupling in Three-Dimensional Trapped-Ion Crystals}

\begin{abstract}
Three-dimensional (3D) crystals offer a route to scale up trapped ion systems for quantum sensing and quantum simulation applications. However, engineering coherent spin-motion couplings and effective spin-spin interactions in large crystals poses technical challenges associated with decoherence and prolonged timescales to generate appreciable entanglement. Here, we explore the possibility to speed up these interactions in 3D crystals via parametric amplification. We derive a general Hamiltonian for the parametric amplification of spin-motion coupling that is applicable to crystals of any dimension in both rf Paul traps and Penning traps. Unlike in lower dimensional crystals, we find that the ability to faithfully (uniformly) amplify the spin-spin interactions in 3D crystals depends on the physical implementation of the spin-motion coupling. We consider the light-shift (LS) gate, and the so-called phase-insensitive and phase-sensitive M\o lmer-S\o rensen (MS) gates, and find that only the latter gate can be faithfully amplified in general 3D crystals. We discuss a situation where non-uniform amplification can be advantageous. We also reconsider the impact of counter-rotating terms on parametric amplification and find that they are not as detrimental as previous studies suggest. 
\end{abstract}

\newcommand{\iisc}{\affiliation{Department of Instrumentation and Applied Physics, Indian Institute of Science, Bangalore 560012, India.}}
\newcommand{\ista}{\affiliation{Institute of Science and Technology Austria, 3400 Klosterneuburg, Austria}}
\newcommand{\bu}{\affiliation{Department of Physics, Boston University, Boston, USA}}
\newcommand{\pitp}{\affiliation{Perimeter Institute for Theoretical Physics, Waterloo, Ontario N2L 2Y5, Canada}}
\newcommand{\jila}{\affiliation{JILA, National Institute of Standards and Technology,and Department of Physics, University of Colorado, Boulder, CO 80309}}
\newcommand{\ctqm}{\affiliation{Center for Theory of Quantum Matter, University of Colorado, Boulder, CO 80309}}
\newcommand{\nist}{\affiliation{National Institute of Standards and Technology, Boulder, CO 80305}}
\newcommand{\NN}[1]{\textcolor{blue}{{\bf NN:} #1}}
\newcommand{\iitm}{\affiliation{Department of Physics, Indian Institute of Technology Madras, Chennai 600036, India}}

\newcommand{\cquicc}{\affiliation{ Center for Quantum Information, Communication and Computing, Indian Institute of Technology Madras, Chennai 600036, India}}

\author{Samarth Hawaldar}
\thanks{These two authors contributed equally to this work.}
\ista
\email{samarth.hawaldar@ist.ac.at}
\author{N. Nikhil}
\thanks{These two authors contributed equally to this work.}
\bu\pitp
\author{Ana Maria Rey}
\jila\ctqm
\author{John J. Bollinger}
\nist
\author{Athreya Shankar}
\email{athreya@physics.iitm.ac.in}
\iitm
\cquicc

\maketitle

\section{Introduction}\label{sec:introduction}

Generating entanglement between two or more qubits on timescales that are short compared to the decoherence times is an essential requirement for many quantum information processing tasks.  Technical limitations typically cause the decoherence effects to become more severe as we try to scale up the number of qubits. Moreover, in systems such as trapped ions, the spin-motion coupling between a single ion and a single motional mode scales down as $1/\sqrt{N}$ with the  number of qubits $N$. Furthermore, the very same lasers that activate the spin-motion coupling are also responsible for elastic and inelastic light scattering processes~\cite{carter2023PRA,moore2023PRA} whose rates per ion remain independent of system size. Hence, it is important to design experimental protocols where the rates of coherent entanglement-generating interactions can be increased without a proportional increase in the rates of background or accompanying decoherence processes.

Recently, parametric amplification (PA) was proposed and experimentally demonstrated as a pathway to increase the coherent spin-motion coupling strength without adding to the decoherence rate~\cite{ge2019PRL,ge2019PRA,burd2021NatPhys,affolter2023PRA}. Very briefly, the technique involves squeezing the motional mode of interest by applying an rf drive to the trap electrodes while simultaneously coupling the motion to the spins using lasers. The combined effect of squeezing and laser driving can be shown to give rise to an effective amplification of the spin-motion coupling strength for fixed laser power, which in turn, keeps the decoherence rate constant. To date, both theory and experiment have considered PA in 1D or 2D trapped-ion crystals, where the set of normal modes often used for quantum information processing is transverse to the spatial extent of the crystal.  

A further scaling up of trapped-ion quantum hardware, especially for quantum simulation and sensing experiments, may rely on utilizing 3D crystals~\cite{wu2021PRA,hawaldar2024PRX,zaris2025JPP}, where no such set of normal modes exist. Hence, the applicability of PA to 3D crystals must be re-examined by relaxing assumptions on the ion crystal geometry that are typically made in prior studies focusing on lower-dimensional crystals. We note that apart from the $1/\sqrt{N}$ scaling with system size, other factors may also lead to weaker spin-motion coupling in 3D crystals. For example, the Lamb-Dicke confinement in 3D crystals may be weaker because of high temperatures or contributions from low-frequency modes, which may necessitate operating with an optical lattice of longer effective wavelength, resulting in weaker spin-motion coupling for a fixed laser intensity. Hence, it becomes important to understand the applicability of PA in 3D crystals.  

In this paper, we model and study PA in a general setting that is broadly applicable to ion crystals of any dimension in both rf Paul traps and Penning traps. A central finding of our work is that in the case of 3D crystals, a faithful amplification of coherent interactions depends on the particular physical realization by which the spin-motion coupling is engineered. This finding is in contrast to previous studies that were restricted to normal modes transverse to the spatial extent of 1D or 2D crystals, where the spin-motion coupling has the same form regardless of the physical realization. On the other hand, we also show that unfaithful amplification can have applications in systems such as bilayer trapped ion crystals~\cite{hawaldar2024PRX}, where it can be used for layer-selective amplification and de-amplification of spin-motion coupling. We also examine the role of counter-rotating terms in limiting the extent of amplification. Our formalism enables us to analytically investigate the impact of counter-rotating terms and shows how to correct for them, thereby making them less detrimental than what has been suggested in previous studies~\cite{ge2019PRL}.  

This paper is organized as follows. In Sec.~\ref{sec:prelims}, we review three physical realizations of spin-motion coupling in trapped ion systems, making no assumptions on the crystal dimensions. Next, in Sec.~\ref{sec:paramp} we derive the Hamiltonian for parametric amplification of spin-motion coupling for a general ion crystal and discuss the 1D, 2D and 3D cases. In Sec.~\ref{sec:bilayer}, we show the utility of unfaithful PA in a large bilayer crystal in a Penning trap. Subsequently, in Sec.~\ref{sec:effect_cr}, we analyze the impact of counter-rotating terms in limiting PA. We conclude with a summary and outlook in Sec.~\ref{sec:conclusion}.

\section{Preliminaries}\label{sec:prelims}
The Hamiltonian of a general 3D ion crystal with $N$ trapped spins (ions) driven by a spin-dependent force (SDF) is given by ($\hbar=1$):
\begin{align}
    \hH &= \sum_{n=1}^{3N} \omega_n \had_n \ha_n + \sum_{j=1}^N \frac{\omega_0}{2} \hs^z_j + \hH_\text{SDF},
\end{align}
where $\omega_n$ are the frequencies of the crystal's normal modes, $\omega_0$ is the frequency of the spin transition, $\hs^z_j$ is the Pauli $Z$ operator for the $j$-th spin, and $\hH_\text{SDF}$ is a spin-dependent force Hamiltonian. $\hH_\text{SDF}$ couples the spin and motional degrees of freedom, giving rise to effective spin-spin interactions mediated by phonons.\\

There are different ways to implement $\hH_\text{SDF}$. We focus on the light-shift (LS) \cite{leibfried2003Nat,britton2012Nat} and M\o lmer-S\o rensen (MS) \cite{sorensen1999PRL,sorensen2000PRA} configurations, both of which use pairs of angled lasers incident on the crystal to induce spin-motion coupling. In an interaction picture taken with respect to the free evolution of the spins, and under the Lamb-Dicke and rotating-wave (RWA) approximations, $\hH$ is generically of the form
\begin{align}
    \hH &= \sum_{n=1}^{3N} \omega_n \had_n \ha_n  + \sum_{j} f_j(t) \hz_j \hs^{\alpha_j}_j.
\end{align}
Here, $\alpha_j$ ($x,y,z$ etc.) is the spin component which couples to the motion along the $z$ direction and $\hz_j$ is the $z$-displacement of the $j$-th ion from its equilibrium position. It is given by 
\begin{align}
    \hz_j &= \sum_n (c_{nj} \ha_n + c_{nj}^* \had_n).
\end{align}
Here, the $c_{nj}$ are mode eigenvector elements, which correspond to the amplitude of ion $j$'s displacement in the $z$ direction due to mode $n$. In general, $c_{nj}$ can be complex, e.g., this is the case in Penning traps, where complex mode eigenvectors arise because of the Lorentz force resulting from the confining magnetic field~\cite{hawaldar2024PRX}.

For a frequency difference $\mu$ between the angled lasers incident on the crystal, the SDF Hamiltonian $\hH$ can be written as
\begin{align}
\label{eqn:GeneralSDF}
    \hH &= \sum_{n=1}^{3N} \omega_n \had_n \ha_n  + \sum_{j} F_j \cos(\mu t - \phi_j) \hz_j \hs^{\alpha_j}_j,
\end{align}
where $F_j$ is the drive strength proportional to the laser intensity used and $\phi_j$ is the relative phase of the SDF drive on ion $j$.

In the so-called `gate' regime of operation~\cite{carter2023PRA}, the frequency $\mu$ is chosen to be close to the frequency $\omega_n$ of a particular mode. Neglecting the far off-resonant contribution of the other modes, the SDF Hamiltonian gives rise to a unitary that factorizes into a spin-motion part and an effective spin-spin part at all times~\cite{wang2013PRA,dylewsky2016PRA}. At specific times $\tau = 2m\pi/(\mu-\omega_n)$, the spin-motion unitary returns to identity and the net effect of the SDF is to engineer a spin-spin interaction. At these times, the effective Hamiltonian governing the dynamics is an Ising interaction of the form  
\begin{align}
    \hH &= \sum_{j,k} J_{jk} \hs^{\alpha_j}_j \hs^{\alpha_k}_k,
\end{align}
where
\begin{align}
    J_{jk} \sim \frac{F_j F_k \Re{c_{nj} c_{nk}^*}}{\mu - \omega_n}.
    \label{eqn:Jij_general}
\end{align}

In the following subsections, we provide a summary of how the SDF Hamiltonian is realized using the LS and MS configurations.

\begin{figure*}
    \centering
    \includegraphics[width=0.8\textwidth]{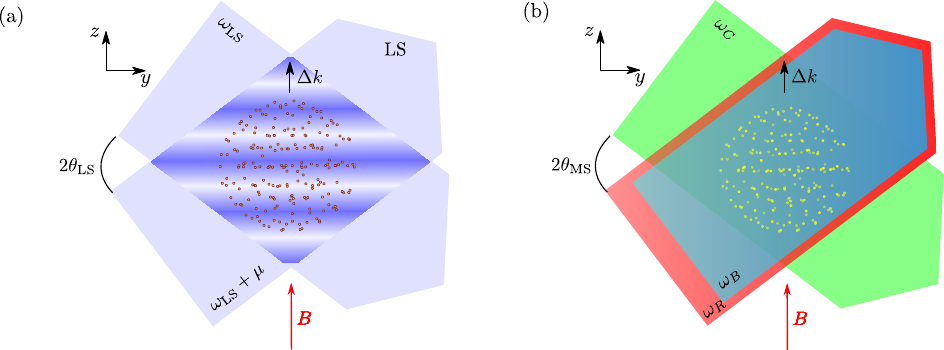}
    \caption{Schematics of the two SDF implementations  for a generic 3D cloud of trapped ions. a) \textbf{Light-shift:} A pair of angled lasers with wavevectors in the YZ plane is incident on the ion cloud. Their interference creates a moving optical lattice along Z, causing spin-motion coupling. b) \textbf{M\o lmer-S\o rensen:} Here, three lasers induce Raman transitions between spin levels. One arm (shown in green) contains the `carrier' beam, and the other arm contains two co-propagating beams called the red/blue sidebands.}
    \label{fig:LS-MS-setups}
\end{figure*}

\subsection{Light-Shift Configuration}
In the LS configuration, as shown in Fig.~\ref{fig:LS-MS-setups}(a), two lasers of frequencies $\omega_{\text{LS}}$, $\omega_{\text{LS}}+\mu$ are incident on the crystal at angles $\pm\theta_{\text{LS}}$ with respect to the crystal plane and off-resonantly couple the spin states to higher electronic states~\cite{britton2012Nat}. For suitable choices of $\theta_{\text{LS}}$ and the beam polarizations, this results in an AC Stark shift with a spatial modulation in the $z$-direction governed by the wavevector $\Delta k \approx 2k\sin(\theta_{\text{LS}})$, where $k = \omega_{\text{LS}}/c$~\cite{sawyer2012PRL}. In an interaction picture with respect to the free evolution of the spins,  this energy shift can be described by the Hamiltonian
\begin{align}
    \hH &= \sum_{n=1}^{3N} \omega_n \had_n \ha_n - \sum_{j} \frac{F_0}{\Delta k} \sin(\mu t - \Delta k(z_{0,j}+\hz_j))  \hs^z_j.
    \label{eqn:ham_ls}
\end{align}
Here, $F_0$ is an effective SDF that is determined by the laser parameters. The Stark shift experienced by each ion depends on the position of the ion, which we have decomposed in Eq.~(\ref{eqn:ham_ls}) into two parts: A constant phase offset $\phi_j = -(\Delta k)z_{0,j}$ that depends on the equilibrium position $z_{0,j}$ of the ion, and a spin-motion coupling term $\Delta k \hz_j$ that depends on the small-amplitude displacement $\hz_j$ of the ion about its equilibrium. In the Lamb-Dicke regime ($\Delta k \sqrt{\ev{\hz_j^2}} \ll 1$) and dropping a motion-independent term under an RWA, the SDF in Eq.~(\ref{eqn:ham_ls}) can be expanded to first order in $\hz_j$ as 
\begin{align}
\label{eqn:LightShiftConfig}
    \hH &\approx \sum_{n=1}^{3N} \omega_n \had_n \ha_n + \sum_{j} F_0 \cos(\mu t + \phi_j)\hz_j  \hs^z_j.
\end{align}
This Hamiltonian matches the general form of an SDF Hamiltonian, shown in Eq.~(\ref{eqn:GeneralSDF}).

\subsection{M\o lmer-S\o rensen Configuration}
The MS configuration utilizes Raman transitions between the spin levels, implemented using a laser configuration such as the one shown in Fig. \ref{fig:LS-MS-setups} (b). Like the LS configuration, there are two arms of lasers incident on the crystal. One arm contains a single beam of frequency $\omega_C$ (denoted the `carrier' beam), and the other carries two co-propagating beams of frequencies $\omega_R$ and $\omega_B$ (denoted the `red' and `blue' sidebands)~\cite{haljan2005PRL,carter2023PRA}. By a suitable choice of laser parameters, the carrier beam is used in conjunction with the red and blue sidebands to implement a pair of two-photon Raman transitions that are also accompanied by changes in motional quanta.  
To second-order in perturbation theory, this interaction can be written in the RWA as
\begin{align}
\label{eqn:MS-Hamiltonian}
    \hat{H} =  &\sum_n \omega_n \hat{a}_n^\dag\hat{a}_n + \sum_{j=1}^N \frac{\omega_0}{2} \hs^z_j \nonumber \\
    + &\sum_j \frac{\Omega_{\rm eff}}{2}\left(\hSig_j^+ e^{i(\mathbf{k_{BC}}\cdot \mathbf{r}-\omega_{BC}t)} + h.c. \right)\nonumber\\
    + &\sum_{j} \frac{\Omega_{\rm eff}}{2}\left(\hSig_j^+ e^{\pm i(\mathbf{k_{RC}}\cdot \mathbf{r}-\omega_{RC}t)} + h.c. \right),
\end{align}
where $\hSig^\pm = \left(\hSig^x \pm i\hSig^y\right)/2$, $\mathbf{k_{BC}} = \mathbf{k_B} - \mathbf{k_C}$, $\omega_{BC} = \omega_B - \omega_C$ (similarly for $\mathbf{k_{RC}}$, $\omega_{RC}$), and 
$\Omega_{\rm eff}$ is an effective two-photon Rabi frequency that is arranged to be equal for the two pairs of Raman transitions.  
 We assume that $\abs{\mathbf{k_{RC}}} \approx 
\abs{\mathbf{k_{BC}}} \approx \Delta k$, and that both difference wavevectors are directed along the positive $z$-axis.\\

\begin{figure}
    \centering
    \includegraphics[width=0.5\textwidth]{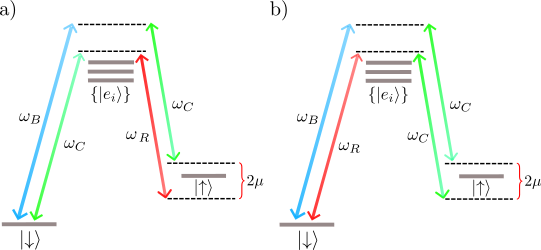}
    \caption{Level diagram illustrating Raman transitions in the phase-insensitive (a) and phase-sensitive (b) configurations of the M\o lmer-S\o rensen gate. The qubit levels $\ket{\uparrow}, \ket{\downarrow}$ are coupled via excited states \{$\ket{e_i}$\}, using the red sideband, the blue sideband, and the carrier beams of frequencies $\omega_R, \omega_B$ and $\omega_C$ respectively.}
    \label{fig:MS-configs}
\end{figure}

The MS gate can be arranged in two distinct configurations depending on the values selected for the three frequency components $\omega_C$, $\omega_B$ and $\omega_R$~\cite{haljan2005PRL, lee2005JOptB}. The choice of configuration determines the sign of the exponent in the third line of Eq.~(\ref{eqn:MS-Hamiltonian}).  These configurations are summarized in Fig.~\ref{fig:MS-configs} and are further discussed below. 

\subsubsection{Phase-insensitive MS}

In this configuration, the carrier beam is used to drive the right (left) leg of a Raman transition in conjunction with the blue (red) sideband, as shown in Fig.~\ref{fig:MS-configs} (a). (Here, the terms `right' and `left' simply refer to the depiction of the transitions in Fig.~\ref{fig:MS-configs} (a).) Accordingly, the sideband frequencies are set to be  $\omega_{B/R} = \omega_C + \mu \pm \omega_0$ and the Hamiltonian~(\ref{eqn:MS-Hamiltonian}) has a negative sign in the exponent in the third line. Moving into an interaction picture with respect to the free evolution of the spins, and expanding to first order in the Lamb-Dicke parameter, the terms with spin-motion coupling are 

\begin{align}
\label{eqn:LambDickeHamiltonian}
    \hH_{\text{spin-motion}} = &\sum_{j}\frac{\Omega_{\rm eff}}{2}(\Delta k) \hz_j \left(i \hSig_j^+ e^{-i(\mu t+\phi_j)} + h.c.\right) \nonumber\\
    - &\sum_{j}\frac{\Omega_{\rm eff}}{2}(\Delta k) \hz_j \left(i \hSig_j^+ e^{i(\mu t+\phi_j)} + h.c.\right).
\end{align}
This Hamiltonian can be simplified to obtain the phase-insensitive MS interaction Hamiltonian as 
\begin{align}
    \label{eqn:pis_MS_ham}
   \hat{H} =  \sum_n \omega_n \hat{a}_n^\dag\hat{a}_n + 
    \sum_{j}\Omega_{\rm eff} \sin(\mu t + \phi_j) \Delta k \hz_j \hSig_j^x,
\end{align}
which has the form of Eq.~(\ref{eqn:GeneralSDF}). The name \textit{phase-insensitive} refers to the property that the same component of spin  couples to the motion for all the spins, independent of the phase $\phi_j$. 

\subsubsection{Phase-sensitive MS}

In this configuration, the carrier is used to drive the right legs of a pair of Raman transitions in conjunction with the blue and red sidebands, as shown in Fig.~\ref{fig:MS-configs}(b). Accordingly, the sideband frequencies are set to be $\omega_{B/R} = \omega_C + \omega_0 \pm \mu$ and the Hamiltonian~ (\ref{eqn:MS-Hamiltonian}) has a positive sign in the exponent in the third line. In the interaction picture associated with the free evolution of the spins, the first-order spin-motion coupling term is now given by 
\begin{align}
    \hH_{\text{spin-motion}} = &\sum_{j}\frac{\Omega_{\rm eff}}{2}(\Delta k) \hz_j \left(i \hSig_j^+ e^{-i(\mu t+\phi_j)} + h.c.\right)\nonumber\\ 
    + &\sum_{j}\frac{\Omega_{\rm eff}}{2}(\Delta k) \hz_j \left(i \hSig_j^+ e^{+i(\mu t-\phi_j)} + h.c.\right).
\end{align}
Hence, the phase-sensitive MS interaction is described by the Hamiltonian 
\begin{align}    
   \hat{H} =  \sum_n \omega_n \hat{a}_n^\dag\hat{a}_n + 
    \sum_{j}\Omega_{\rm eff} \cos(\mu t) \Delta k \hz_j \hSig_j^{\phi_j}.
\end{align}
Here, $\hSig_j^{\phi_j} = \sin(\phi_j)\hSig^x_j - \cos(\phi_j)\hSig^y_j$ denotes the spin component in a direction angled at $\phi_j$ to the $-Y$ axis in the Bloch sphere's $XY$ plane. This Hamiltonian differs from (\ref{eqn:pis_MS_ham}) in that the component of the spin coupling to the motion depends on $\phi_j$, and hence it is called the \textit{phase-sensitive} configuration. 

For state initialization and readout performed using the same Raman lasers as that used for the MS gate, this spin phase can be absorbed into the definition of a local frame for each spin that can be consistently used throughout the operation i.e. we can replace $\hSig_j^{\phi_j}$ with $\hSig_j^{x}$ under the understanding that the $x$ axis is defined in a local, ion-specific frame that is valid when using the same Raman lasers for all operations. 
Hence, for the remainder of the article, we write the phase-sensitive MS Hamiltonian as
\begin{align}
\label{eqn:ps_ms_ham}
   \hat{H} =  \sum_n \omega_n \hat{a}_n^\dag\hat{a}_n + 
    \sum_{j}\Omega_{\rm eff} \cos(\mu t) \Delta k \hz_j \hSig_j^x.
\end{align}

\section{Parametric Amplification}\label{sec:paramp}
The implementation of the SDF using laser-induced spin-motion coupling additionally leads to scattering processes at a rate that is proportional to the laser intensity. The associated decoherence limits the fidelity of coherent spin-motion and phonon-mediated spin-spin interactions. One approach to mitigate the adverse impact of decoherence is to use the technique of parametric amplification to boost the spin-motion coupling strength for a fixed laser intensity. In this section, we will analyze the net effect of simultaneous PA and SDF driving on the spin-motion coupling strength, and then discuss the effect of PA on the spin-motion coupling in 1D, 2D and 3D trapped ion crystals.


\subsection{Analysis of simultaneous PA + SDF driving}

Parametric amplification (PA) is the process of non-linearly amplifying the amplitude of a quantum harmonic oscillator by subjecting it to a squeezing drive. In the context of trapped-ion spin-motion coupling, applying PA on a motional mode while simultaneously coupling it to the spins via the SDF leads to an effective amplification of the SDF~\cite{ge2019PRL, ge2019PRA}. In practice, PA is achieved by adding a small-amplitude modulation to the harmonic trapping potential at twice the SDF beat frequency $\mu$ [see Eq.~(\ref{eqn:H_PA_gen})]. The full expression for the trapping potential associated with the modulation satisfies Laplace's equation and has the form~\cite{heinzen1990PRA}
\begin{equation}
    V_\text{PA}(t) = V_p \cos(2\mu t-\theta) \sum_j\bigg(z_j^2 - \frac{1}{2}\rho_j^2 \bigg),
\end{equation}
where $V_p$ is a voltage amplitude, $\rho_j = \sqrt{x_j^2 + y_j^2}$ is the radial distance of ion $j$ from the origin, and $\theta$ is the modulation phase. As shown below, this modulation generates a squeezing Hamiltonian $\hHp$ which amplifies the effective spin-motion coupling strength when it is simultaneously applied with the SDF.

Taking the LS gate Hamiltonian as an example, the total Hamiltonian including the parametric drive for the most general ion crystal geometry is given by 
\begin{align}
 \hH &= \hH_\text{0} + \hH_{\text{SDF}} + \hHp,\label{eqn:H_tot}\\
 \hH_\text{0} &= \sum_n \omega_n \had_n \ha_n,\\
 \hH_{\text{SDF}} &= \sum_j F_0 \cos(\mu t + \phi_j) \hz_j \hSig_j^z,\label{eqn:H_SDF_gen}\\
 \hHp &= \Omega_p \cos(2\mu t-\theta)\nonumber\\
 &\sum_j \left((z_{0,j} + \hz_j)^2 - \frac{1}{2}(x_{0,j}+\hx_j)^2 - \frac{1}{2}(y_{0,j}+\hy_j)^2\right),\label{eqn:H_PA_gen}
\end{align}
where $\Omega_p \propto V_p$ is the effective drive that depends on the applied modulation voltage and the trap size. This Hamiltonian is in an interaction picture taken with respect to the free evolution of the spins. Furthermore, we note that we have dropped a motion-independent ac Stark shift term proportional to $\sin(\mu t + \phi_j)$ in $\hH_{\rm SDF}$ under a RWA.

In the case of ion crystals in Penning traps, the crystal is typically rotating in the lab frame about the $z$ axis at a precisely controlled frequency. We note that $\hHp$ is invariant under a rotating frame transformation and the equilibrium positions and displacements entering $\hHp$ can be taken to represent quantities in the frame rotating at the crystal rotation frequency, which is commonly used in modeling Penning trap experiments~\cite{wang2013PRA,shankar2020PRA}. 

\subsubsection{Normal mode decomposition}

We can decompose the ion displacements in terms of the quantized normal modes of the ion crystal. For any ion $j$, we can write its displacement $\hat{\alpha}_j$ (for $\alpha \in \{x,y,z\}$) using the ladder operators of the normal modes as 
\begin{align}
\label{eqn:generalModeExpansion}
 \hat{\alpha}_j = \sum_n l_{n}\left(u_{nj}^\alpha \ha_n + u_{nj}^{\alpha*} \had_n\right),
\end{align}
where $l_n$ is a normalized zero-point fluctuation of the mode $n$, and $u_{nj}^\alpha$ are complex numbers representing the components of the mode $n$ along the direction $\alpha$ for the ion $j$, with the constraint that  
\begin{align}
 \sum_{\alpha,j} u_{nj}^\alpha u_{nj}^{\alpha*} = 1.
\end{align}
In writing Eq.~(\ref{eqn:generalModeExpansion}), we have allowed for the possibility of complex normal mode eigenvectors, which arise in Penning traps because of the confining magnetic field~\cite{hawaldar2024PRX}. In the special case of 2D planar crystals in Penning traps, the in-plane and out-of-plane (parallel to the magnetic field and taken as the $z$ direction) motion decouple, and the latter can be decomposed purely in terms of real eigenvectors~\cite{wang2013PRA}. However, in the general case of 3D crystals, motion parallel and perpendicular to the magnetic field no longer decouple and we require complex eigenvectors to decompose the $z$-direction motion. Furthermore, the Lorentz force also modifies the orthogonality condition of the complex normal modes in Penning traps~\cite{shankar2020PRA}. Hence, our formalism \emph{does not} assume the usual mode orthogonality condition, which for two modes $n,m$ is given by  $\sum_{\alpha,j} u_{nj}^\alpha u_{mj}^{\alpha*}=\delta_{n,m}$. \\

We first express the Hamiltonian in the absence of the parametric drive using the normal modes. Transforming the modes into a frame rotating at the frequency $\mu$ of the SDF drive, we obtain 
\begin{multline}
\label{eqn:RWA_ODF}
 \hH_\text{0} + \hH_{\text{SDF}} = -\sum_n \delta_n \had_n \ha_n \\
 + \sum_{j,n} f_n \left(u_{nj}^z \ha_n e^{i\phi_j} + u_{nj}^{z*} \had_n e^{-i\phi_j}\right) \hSig_j^z,
\end{multline}
where $\delta_n = \mu-\omega_n$, $f_n = F_0 l_n/2$ and we have neglected the terms rotating at a frequency $2\mu$. The full Hamiltonian without this RWA is shown in Appendix~\ref{app:CR_full_LS} and the effects of the fast-rotating terms are discussed in Sec. \ref{sec:effect_cr}. 

\begin{figure*}[!ht]
    \centering
    \includegraphics[width=0.9\linewidth]{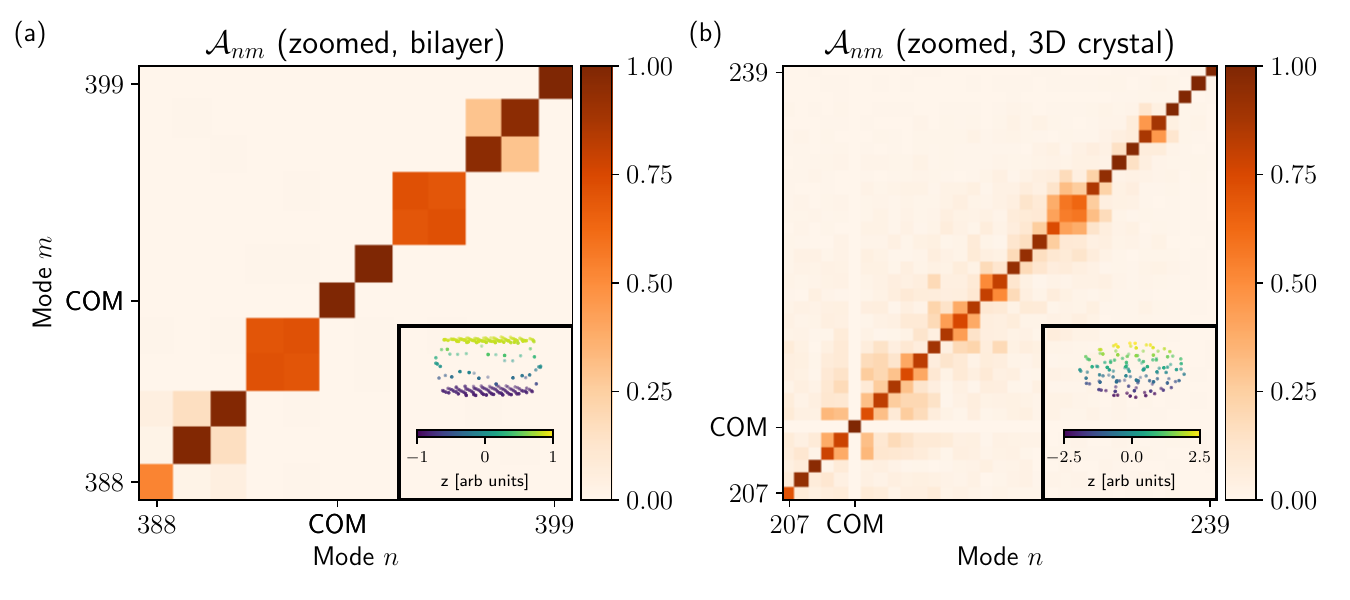}
    \caption{$\sA_{nm}$ matrix for the highest few drumhead modes of (a) a bilayer crystal of 200 $^9$Be$^+$ ions in a Penning trap with an anharmonic trapping potential (see Fig.~3 and Fig.~4 of Ref.~\cite{hawaldar2024PRX} for the list of trap parameters leading to the formation of this crystal), and (b) a 3D crystal of 120 $^9$Be$^+$ ions in a Penning trap with magnetic field of $4.4588\,$T, axial trapping frequency of $1.62\,$MHz, rotating wall frequency of $400\,$kHz, and a dimensionless rotating wall strength of $0.015$. The inset in each plot shows the crystal for which the $\sA_{nm}$ values were calculated. The definitions of various trap parameters can be found in Ref.~\cite{hawaldar2024PRX}.}
    \label{fig:Anm_3d}
\end{figure*}

To express $\hHp$ in terms of the normal modes, we first drop the constant terms of $\hHp$ and write it as 
\begin{align}
 \hHp &= \Omega_p \cos(2\mu t-\theta)\times \nonumber\\
 &\sum_j \left(\hz_j^2 - \frac{\hx_j^2 + \hy_j^2}{2} + 2z_{0,j}\hz_j - x_{0,j}\hx_j - y_{0,j}\hy_j \right).
\end{align}
We now define
\begin{align}
\label{eqn:g_nm,A_nm defn}
 g_{nm} &= \Omega_p l_n l_m, \nonumber\\
 A_{nm} &= \sum_j \left(u_{nj}^z u_{mj}^z - \frac{u_{nj}^x u_{mj}^x + u_{nj}^y u_{mj}^y}{2} \right).
\end{align}
These quantities are helpful in compactly expressing $\hHp$ in terms of the mode creation and annihilation operators. Here, $g_{nm}$ can be understood as the strength of an effective non-linear drive involving modes $n$ and $m$, and $A_{nm}$ can be interpreted as a parameter quantifying an overlap between the modes and the PA drive. For 1D and 2D crystals, the target normal modes for spin-motion coupling are usually taken to be transverse to the spatial extent of the crystal. Such modes are typically decoupled from the motion along the crystal extent and satisfy the typical orthogonality condition. As a result, $A_{nm}=\delta_{nm}$ is a valid assumption for analyzing PA in 1D and 2D crystals. However, more generally, $A_{nm}\neq \delta_{nm}$ for all pairs of modes $m,n$ in 3D crystals. As an example, in Fig.~\ref{fig:Anm_3d}, we plot $A_{nm}$ over a select range of drumhead modes (modes predominantly along the $z$ direction) in bilayer and 3D spheroidal crystals formed in Penning traps. We observe that $A_{nm}$ is typically non-diagonal, although specific modes such as the centre-of-mass (COM) mode do not couple to other modes.  In addition, we note that as a mode eigenvector is only uniquely defined up to a global phase, the phase of the complex number $A_{nm}$ depends on the global phases chosen for the $3N$ normal modes. However, observable quantities are independent of this choice.

Now, moving into a frame rotating at the SDF frequency $\mu$, writing the position operators in terms of normal modes, and keeping only the time-independent terms under an RWA (derived in Appendix \ref{app:CR_full_LS}), we obtain
\begin{align}
\label{eqn:RWA_PA}
 \hHp &= \sum_n \frac{g_{nn}}{2} \left(\ha[2]_n e^{-i\theta} A_{nn} + \had[2]_n e^{i\theta} A_{nn}^* \right)\nonumber\allowdisplaybreaks\\
 &+\sum_{n\ne m} \frac{g_{nm}}{2} \left(\ha_n \ha_m e^{-i\theta} A_{nm} + \had_n \had_m e^{i\theta} A_{nm}^* \right).
\end{align}

\subsubsection{Bogoliubov transformation}
\label{sec:bgb}
Since $\hHp$ is now a time-independent Hamiltonian that is quadratic in the ladder operators, we can transform $\hH_{0} + \hHp$ into a set of free, non-interacting modes using a Bogoliubov transformation. In principle, the transformation should account for the mode mixing terms ($n\neq m$ terms) in Eq.~(\ref{eqn:RWA_PA}). However, in practice, we are typically interested in coupling near-resonantly with specific modes, such as the COM mode, that have negligible coupling $A_{nm}$ to other modes. Hence, we neglect the mode-mixing terms and introduce new mode operators $\hb_n,\hbd_n$ related to $\ha_n,\had_n$ as 
\begin{align}\label{eqn:a_to_b_modes}
 \ha_n &= \hb_n \cosh(r_n) - \hbd_n \sinh(r_n) e^{i\varphi_n},
\end{align}
where $r_n$ is a real parameter that quantifies the amount of squeezing and $\varphi_n$ specifies the quadrature that is squeezed.
Expressing $\hHp$ in terms of $\hb_n$ and $\hbd_n$, we can write the coefficients of the two-phonon annihilation term ($\hb[2]_n$) for the $n$-th mode as
\begin{multline}
\label{eqn:b^2coeff}
 \delta_n \sinh(r_n) \cosh(r_n) e^{-i\varphi_n} \\+ \frac{g_{nn}}{2} \left(A_{nn} \cosh[2](r_n) e^{-i\theta} + A_{nn}^{*} \sinh[2](r_n) e^{i(\theta - 2\varphi_n)} \right).
\end{multline}
Similarly, the coefficient of the $\hbd_n\hb_n$ term is
\begin{multline}
 -\delta_n (\cosh[2](r_n) + \sinh[2](r_n)) \\- 2g_{nn} \cosh(r_n)\sinh(r_n) \Re{A_{nn} e^{-i(\theta - \varphi_n)}}.
\end{multline}
To diagonalize $\hHp$, we can now solve for $r_n$ and $\varphi_n$ such that the coefficient of $\hb[2]_n$ vanishes. 

Now, using the transformation~(\ref{eqn:a_to_b_modes}) on the Hamiltonian $H_{\rm SDF}$ results in an effective spin-motion coupling with the new set of modes that has the form
\begin{align}
\label{eqn:ODF_b_modes}
 \hH_{\rm SDF} = \sum_{j,n} \left( f_n' \hb_n + f_n'^{*} \hbd_n \right) \hSig_j^z,
\end{align}    
where the new, modified coupling strengths are given by 
\begin{align}
 f_n' = f_n\left(u_{nj}^z e^{i\phi_j} \cosh(r_n) - u_{nj}^{z*} e^{-i\phi_j} \sinh(r_n) e^{-i\varphi_n}\right).
\end{align}

\subsection{Parametric Amplification of Transverse Modes in 1D and 2D crystals}

In this section, we review the effect of a parametric drive in 1D and 2D crystals, where the objective is to amplify the spin-motion coupling between the ions and one or more normal modes that are transverse to the spatial extent of the crystal~\cite{ge2019PRL,ge2019PRA}. Subsequently, we will generalize the analysis to 3D crystals in the next section. 

In this setting, all the ions lie in a single plane perpendicular to the direction of the optical lattice generated by the SDF lasers, taken here to be the $z$ direction. Hence, all ions can be taken to lie in the $z=0$ plane and the phase offset $\phi_j=0$ $\forall j$. Furthermore, the normal modes in the $z$-direction are decoupled from motion in the $x$--$y$ plane. In rf Paul traps, this decoupling ensures that the expression~(\ref{eqn:g_nm,A_nm defn}) for $A_{nm}$ reduces to the orthogonality relation of the transverse modes, which makes $A_{nm} = \delta_{nm}$.  In Penning traps, an additional subtlety must be considered in general 3D crystals, namely that the mode orthogonality relations are modified due to the Lorentz force associated with the magnetic field~\cite{shankar2020PRA}. However, in 2D Penning trap crystals, the decoupling of transverse modes from the in-plane modes ensures that there is no effect of the magnetic field on the orthogonality conditions for the transverse modes, and hence $A_{nm}=\delta_{nm}$ even for this case. Moreover, in both rf Paul traps and Penning traps, the transverse mode eigenvectors can be taken to be purely real. 

Hence in this setting, the Bogoliubov transformation neglecting the mode mixing terms is exact. Concretely, for the LS gate considered above, we set the coefficient of $\hb[2]_n$ in (\ref{eqn:b^2coeff}) to zero and obtain 
\begin{align}
 \varphi_n = \theta, \ \ e^{-r_n} = \left(\frac{\delta_n + g_{nn}}{\delta_n - g_{nn}}\right)^{1/4}.
 \label{eqn:bgb_sol}
\end{align}
We note that $r_n$ as obtained from Eq.~(\ref{eqn:bgb_sol}) is a negative real number. This property arises because of the minus sign convention adopted in the Bogoliubov transformation in Eq.~(\ref{eqn:a_to_b_modes}).

Substituting these values, the full Hamiltonian~(\ref{eqn:H_tot}) in terms of the new modes $\hb_n,\hbd_n$ is given by 
\begin{align}
 \hH &= \sum_n \left(-\delta_n' \hbd_n\hb_n + f_n  \left( \hb_n S_n + \hbd_n S_n^* \right) \sum_j u_{nj}^z \hSig_j^z \right),\label{eqn:ham_full_bgb}\\
 \delta_n' &= \sqrt{\delta_n^2 - g_{nn}^2},\\
 S_n &= \cosh(r_n) - \sinh(r_n) e^{-i\theta}.
\end{align}

Equation~(\ref{eqn:ham_full_bgb}) shows that the effect of the parametric drive is to scale the spin-motion coupling strength by a factor $\abs{S_n}$ for each mode. Importantly, the quantity $S_n$ is the same for all ions, and we refer to this feature as \textit{uniform} or \textit{faithful} amplification since there is no distortion of the relative coupling strengths inherent in the original SDF Hamiltonian. Notably, the above analysis carries through in exactly the same manner for all three gates considered in this work, namely the LS, the phase-sensitive MS and the phase-insensitive MS gates, with the change that $\hSig_j^z\to\hSig_j^x$ in Eq.~(\ref{eqn:ham_full_bgb}) for the latter two gates. Hence, all three gates can be  faithfully amplified using a parametric drive, provided the normal modes used are transverse to the spatial extent of the crystal.

\subsection{Parametric Amplification in 3D}
We now turn to the case of 3D crystals, which is the focus of this work. In general, the effect of a parametric drive on the spin-motion coupling in 3D crystals can be richer than in 1D and 2D crystals, primarily for three reasons. First, since the crystal has spatial extent in all three dimensions, there is no set of transverse modes and hence $\phi_j$ cannot be set to $0$ for all ions, introducing spatial inhomogeneity in the phase offset of the SDF across the crystal. This property is a feature of the SDF itself, and is independent of whether or not a parametric drive is applied. Second, the modes in the direction of interest, taken again to be the $z$-direction, do not in general decouple from motion in the $x-y$ plane, thereby making $A_{nm}$ inequivalent to an orthonormality condition. Hence, in general $A_{nm} \neq 0$ for $n\neq m$, and $A_{nn} \neq 1$. Third, 3D crystals in Penning traps support chiral modes with complex eigenvectors, even in mode branches that are primarily along the $z$ direction~\cite{hawaldar2024PRX}. 

Assuming that the modes of interest do not mix or mix very weakly, such as the COM mode (see Fig.~\ref{fig:Anm_3d}), we neglect the mixing terms $A_{nm}$ and consider the modified modes described by Eq. (\ref{eqn:a_to_b_modes}). Requiring that the coefficient of $\hb[2]_n$ vanish leads to the solution 
\begin{align}
 \varphi_n = \theta, \ \ e^{-r_n} = \left(\frac{\delta_n + g_{nn}\sA_{nn}}{\delta_n - g_{nn}\sA_{nn}}\right)^{1/4}.
 \label{eq:exp_rn_3d}
\end{align}
In terms of the new modes $\hb_n,\hbd_n$, the full Hamiltonian is now given by
\begin{align}
 \hH &= \sum_n \left(-\delta_n' \hbd_n\hb_n + f_n \sum_j (v_{nj}^z \hb_n + v_{nj}^{z*} \hbd_n)\hSig_j^z\right),\\
 \delta_n' &= \sqrt{\delta_n^2 - g_{nn}^2 \sA_{nn}^2},\label{eqn:detuning_eff_3d}\\
 v_{nj}^z &= u_{nj}^z e^{i\phi_j} \cosh(r_n) - u_{nj}^{z*} e^{-i\phi_j} \sinh(r_n)e^{-i\theta}.
\end{align}
We note that $\hSig_j^z\to\hSig_j^x$ for the MS gates. From these expressions, one can see that the Hamiltonian retains the form of an SDF, but with modified mode amplitudes ($u_{nj}^z \rightarrow v_{nj}^z$). Equivalently, we can write $v_{nj}^z = S_{nj} u_{nj}^z$ where
\begin{align}
\label{eqn:snj_3d}
 S_{nj} = \cosh(r_n) - e^{-i(\theta + 2\phi_j + 2\arg{u_{nj}^z})} \sinh(r_n)
\end{align}
is the scaling factor for the spin-motion coupling strength, now dependent on both the mode $n$ \emph{and} the ion $j$. 

In 1D and 2D crystals, $\phi_j=0$ $\forall j$ and the eigenvectors are real, which makes $S_{nj}$ independent of ion index $j$. In the case of 3D crystals, $\phi_j$ is dependent on $j$ and the mode eigenvector can be complex, making $S_{nj}$ dependent on $j$, in general. This inhomogeneity in the scaling factor prevents a parametric drive from faithfully amplifying the SDF in the most general setting. 

However, in the specific situation that the phase-sensitive MS configuration is used to couple the ions to (predominantly) real normal modes, such as the COM mode, $\phi_j$ can be absorbed into the definition of a local reference frame for the spins provided the same set of Raman lasers is used to perform global single-qubit rotations as well as implement the MS gate. Hence, $\phi_j$ can be set to $0$. Further, for \textit{real} eigenvectors, $e^{-2i\arg{u_{nj}^z}} = 1$ i.e. $S_{nj}$ can be made independent of $j$, resulting in faithful amplification. Hence, using the phase-sensitive MS configuration and an appropriate choice of mode, the challenges presented by a 3D crystal can be circumvented. 

\begin{figure*}
    \centering
    \includegraphics[width=\linewidth]{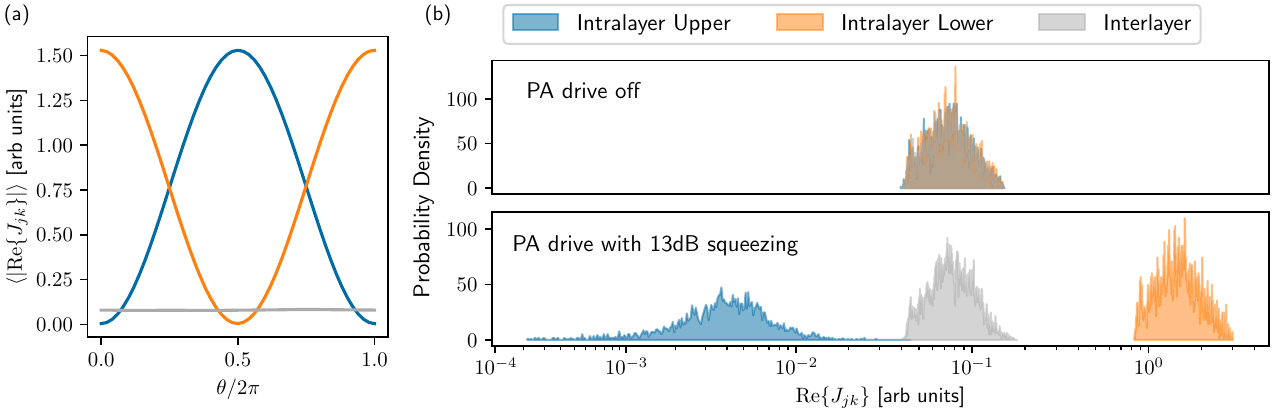}
    \caption{Layer Selective Amplification of the COM mode of a bilayer crystal with $10\log_{10} (e^{-2r_n}) = 13$ dB of squeezing, obtained by setting $\delta_n' = 2\pi\times 1$ kHz and $g=2\pi\times10$ kHz. (a) Spin-spin interaction strength, averaged over all ion pairs with the two ions in the same (blue, orange) or different layers (gray), as a function of relative phase $\theta$ between the PA drive and the SDF for an interlayer phase $\Phi=\pi/2$, (b) Histogram of spin-spin interaction strengths between all the ion pairs in the absence (top) or presence (bottom) of a parametric drive at a relative phase $\theta=0$ with respect to the SDF. In both panels (a) and (b), the ``scaffolding'' ions, i.e. ions at the extremities of the crystal and not belonging to any layer, have not been considered [see inset of Fig.~\ref{fig:Anm_3d} (a)].}
    \label{fig:Bilayer_squeezing}
\end{figure*}

\section{Unfaithful parametric amplification: An application \label{sec:bilayer}}

Although certain SDF configurations are not amenable to faithful amplification under a parametric drive, unfaithful amplification can become a useful resource for controlling interactions in certain settings. In this section, we demonstrate one such application of unfaithful parametric amplification to the case of bilayer trapped ion crystals in Penning traps. 

\subsubsection{Layer-selective amplification for Bilayer Crystals}

Bilayer crystals, such as the one shown in the inset of Fig.~\ref{fig:Anm_3d} (a), are an attractive platform for quantum simulation experiments~\cite{hawaldar2024PRX}. In particular, their applications can be versatile if the strengths of the interlayer and intralayer spin-spin interactions can be dynamically tuned. In Ref.~\cite{hawaldar2024PRX}, a method was proposed to tune the relative strength of interlayer versus intralayer spin-spin interactions using two SDF drives on two separate modes. Here, we show how to achieve a complementary capability: We propose a method to tune the relative strength of the intralayer interactions in the upper and lower layers of the crystal, using only one mode in conjunction with unfaithful parametric amplification of a global LS or a phase-insensitive MS gate. 

To illustrate the central idea, we consider an ideal bilayer crystal in a traveling-wave optical lattice, such that $\phi_j=0$ for all ions in the top layer and $\phi_j=\Phi$ for all ions in the bottom layer. We assume that the SDF generated by the optical lattice couples primarily to the COM mode in the direction transverse to the plane of the layers, taken to be the $z$ direction. As the COM mode is predominantly real, $\arg{u_{nj}^z}\approx0$, and $S_{nj}$ [Eq.~(\ref{eqn:snj_3d})] is constant for all ions in a single layer, with
\begin{align}
\label{eqn:blr_sn}
 \abs{S_{n, \text{Top}}} &=\sqrt{\cosh(2r_n) + \cos(\theta) \sinh(2r_n)},\nonumber\\
 \abs{S_{n, \text{Bottom}}} &=\sqrt{\cosh(2r_n) + \cos(\theta-2\Phi) \sinh(2r_n)}.
\end{align}

In the so-called gate regime of operation, spin-motion coupling is mediated by a single mode $n$, leading to effective Ising type spin-spin interactions of the form given in Eq.~(\ref{eqn:Jij_general}) at the decoupling times~\cite{ge2019PRL}. In the presence of a parametric drive, the resulting amplification leads to a scaling of the effective spin-spin coupling strengths, such that    
\begin{equation}
    J_{jk} \to S_{nj}S_{nk} J_{jk}.
\end{equation}

In Fig.~\ref{fig:Bilayer_squeezing}, we compute the scaled spin-spin coefficients for the numerically simulated bilayer crystal shown in the inset of Fig.~\ref{fig:Anm_3d} (a). We only plot $\Re{J_{jk}}$ since the imaginary parts vanish when taking the sum over all ion pairs in the Ising Hamiltonian. In panel (a), we compute the average value of this quantity over all pairs of ions, where the two ions belong to the upper (blue), lower (orange) or different (gray) layers. As the parametric drive phase $\theta$ relative to the SDF is varied, we observe that the strength of intralayer interactions in the two layers can be continuously tuned over a range of values, while the interlayer interactions remain constant.  In panel (b), we consider the case when $\Phi=\pi/2$ and $\theta=0$, and compare the histograms of the intralayer and interlayer spin-spin interaction strengths in the absence and presence of the parametric drive. For this choice of parameters, the histograms show that almost all pairs of ions in the upper (lower) layer have significantly suppressed (enhanced) interactions, while the interlayer interactions for all pairs are unaffected by the parametric drive. The histograms show that the suppression and enhancement of intralayer interactions in the upper and lower layers are robust to variations in the $z$ positions of ions in realistic bilayer crystals, which we neglected while obtaining the scale factors in Eq.~(\ref{eqn:blr_sn}). 

\section{Effect of Counter-Rotating terms}\label{sec:effect_cr}

In Ref.~\cite{ge2019PRL}, the impact of several experimental imperfections and modeling assumptions on PA was considered in detail. While most of these analyses carry through directly to the 3D crystal case, here, we revisit the impact of counter-rotating terms that were neglected in our analysis in previous sections. Our analysis provides further insight into the physical implications of the counter-rotating terms and also suggests that they may not be as detrimental as reported in previous studies.

In this section, we will analyse the effect of counter-rotating (CR) terms  for the phase-sensitive MS configuration of the SDF. The counter-rotating terms in the other configurations contribute in a similar manner and the insights gained from this analysis can be generalized across all the three configurations considered in this paper.

As we show below, in the frame rotating with the frequency $\mu$ of the SDF drive, the full Hamiltonian of our system under a parametric drive can be written in the form
\begin{align}
 \hH &= \sum_{l=-\infty}^\infty \sH_l e^{il\mu t}.
\end{align}
For \textit{time-dependent periodic} Hamiltonians of the above form, the Floquet theorem guarantees that the evolution of the whole system at times $\Delta t=2n\pi/\mu$ can be exactly described using a \textit{time-independent} Hamiltonian called the Floquet Hamiltonian \cite{Bukov2015AdvPhys}. 

As the frequency $\mu$ is much larger than all the other frequencies in our problem, we can approximately obtain the Floquet Hamiltonian $\sH_F$ using a high-frequency expansion in powers of $1/\mu$~\cite{Bukov2015AdvPhys}, i.e.
\begin{align}
\label{eqn:floquet_ham}
 \sH_F = \sH_0 + \sum_{l\ne 0} \frac{[\sH_l, \sH_{-l}]}{l \hbar \mu } + \sO\left(\frac{1}{(\hbar\mu)^2}\right).
\end{align}

We now proceed to write down the counter-rotating terms in the present setting. We group the counter-rotating terms into different Hamiltonians depending on whether they come from the SDF or the parametric drive. Defining $f_n=\Omega_{\text{eff}}\Delta k l_n/2$, we obtain the counter-rotating contributions to be (see Appendix~\ref{app:CR_full_PSMS}) 
\begin{align}
\label{eqn:CR-terms-SDF}
	\hH_\text{CR, SDF} &= \sum_j \frac{\Omega_\text{eff}}{2}\;\hSig_j^y\; (e^{i\mu t}+ e^{-i\mu t}) \nonumber\\
	&+ \sum_{j,n} f_n \left(u_{nj}^{z*} \had_n e^{i(2\mu t - \phi_j)} + \HC\right) \hSig_j^x,\allowdisplaybreaks\\
\label{eqn:CR-terms-PA}
	\hH_\text{CR, PA} &= \sum_{n,m} \frac{g_{nm}}{2} \left(\ha_n\ha_m e^{-i(4\mu t + \theta)} A_{nm} + \HC \right)\nonumber\\
	&+ \sum_{n,m} g_{nm} (e^{i(2\mu t - \theta)}B_{nm}^* + \HC )\had_n \ha_m \nonumber\\
	&+ \sum_{j,n} K_{jn} \left(\ha_n e^{-i\theta}\left(e^{i\mu t} + e^{-3i\mu t}\right) + \HC \right).
    \end{align}

Here, $B_{nm}$ is a quantity similar to $A_{nm}$, and is given by 
    \begin{align}
	B_{nm} &= \sum_j \left(u_{nj}^{z*} u_{mj}^z - \frac{u_{nj}^{x*} u_{mj}^x + u_{nj}^{y*} u_{mj}^y}{2} \right),
\end{align}
while $K_{jn}$ is a mode and ion-dependent complex number whose explicit form we do not provide here; it is not important for further calculations since we find that the odd harmonic terms do not mix with other terms at leading order, and only contribute a constant shift to the Hamiltonian.

Using Eq.~(\ref{eqn:floquet_ham}) and considering only $\hH_\text{CR, SDF}$, we find the correction terms (denoted using the script symbol $\sH$) 
\begin{align}
	\sH_\text{CR, SDF} &= - \sum_{j\ne k} \left[\sum_n \frac{f_n^2 u_{nj}^{z*} u_{nk}^{z}e^{i(\phi_k-\phi_j)}}{2\mu} \right] \hSig_j^x \hSig_k^x.
\end{align}
Similarly, considering only $\hH_\text{CR, PA}$, we evaluate the Floquet correction term 
\begin{align}
	\sH_\text{CR, PA} &= \sum_{n,n',m,m'} \frac{g_{nm} g_{n'm'} A_{nm}^* A_{n'm'}}{16\mu} [\had_n \had_m, \ha_{n'}\ha_{m'}]\nonumber\\
	&+ \sum_{n,n',m,m'} \frac{g_{nm} g_{n'm'} B_{nm} B_{n'm'}}{2\mu} [\had_n \ha_m, \had_{n'}\ha_{m'}]\allowdisplaybreaks\nonumber\\
	&= -\sum_{n,n'} \frac{\had_n \ha_{n'}}{4\mu} \left(\sum_m g_{nm} g_{n'm} A_{nm}^* A_{n'm}\right). \label{eqn:CorrPA}
\end{align}
Here, the contribution of the $2\mu$-term becomes zero due to its anti-symmetry under an exchange of indices $n~\leftrightarrow~m$. We observe that the CR terms coming from the  SDF contribute an extra XX-coupling between the spins. On the other hand, for the CR terms coming from the parametric drive, the diagonal ($n=n'$) terms contribute (Bloch-Siegert-like) frequency shifts and the off-diagonal ($n\neq n'$) terms lead to beam-splitter-type couplings between modes. Finally, we also observe that the $2\mu$-terms in the SDF and PA drives can mix together to give another correction of the form
\begin{align}
	\sH_\text{CR, SDF-PA} &= - \sum_{j,n,m} \left(g_{nm} f_{n} u_{nj}^z B_{nm}^* e^{i\phi_j} \ha_m +\HC\right) \frac{\hSig_j^x}{2\mu},
\end{align}
which just amounts to an extra SDF-type term. Finally, taking into account all these contributions, we can write the effective Floquet Hamiltonian as
\begin{equation}
    \sH_F = \sH_0 + \sH_\text{CR, SDF} + \sH_\text{CR, PA} + \sH_\text{CR, SDF-PA},
\end{equation}
where $\sH_0$ is the co-rotating contribution of the full Hamiltonian in a frame rotating with the SDF drive (the expression for $\sH_0$ for the phase-sensitive MS gate is given by Eq.~(\ref{eqn:H_psms_RWA_full})).

We see that only the CR terms from the parametric drive cause shifts in the mode frequencies, and hence enter the Bogoliubov transformation in Sec.~\ref{sec:bgb} through the detunings $\delta_n$. Neglecting the cross-terms in Eq.~(\ref{eqn:CorrPA}) and denoting $G_n^2\equiv\sum_m g^2_{nm} \sA_{nm}^2$, we find that the detuning $\delta_n$ must be shifted to  
\begin{align}
        \label{eqn:deltan_corr}
	\delta_n \longrightarrow \delta_n - \frac{G_n^2}{4\mu}
\end{align}
in order to compensate for the frequency shift arising from the CR terms.  Provided the detuning is suitably shifted, the gain expression is still given by Eq.~(\ref{eq:exp_rn_3d}). On the other hand, if the detuning is not adjusted to compensate for the frequency shift, the effective amplification of the spin-motion coupling is instead given by 
\begin{align}
    e^{-r_n} = \left(\frac{\delta_n + G_n^2/4\mu + g_{nn}\sA_{nn}}{\delta_n + G_n^2/4\mu - g_{nn}\sA_{nn}}\right)^{1/4}.
    \label{eq:exp_rn_3d_cr}
\end{align}
We note that, for most modes of interest, $\sA_{nm}\approx \delta_{nm}$ and hence we assume $G_n^2 = g_{nn}^2$ in our discussion below.

\begin{figure}[h]
    \centering
    \includegraphics[width=0.8\linewidth]{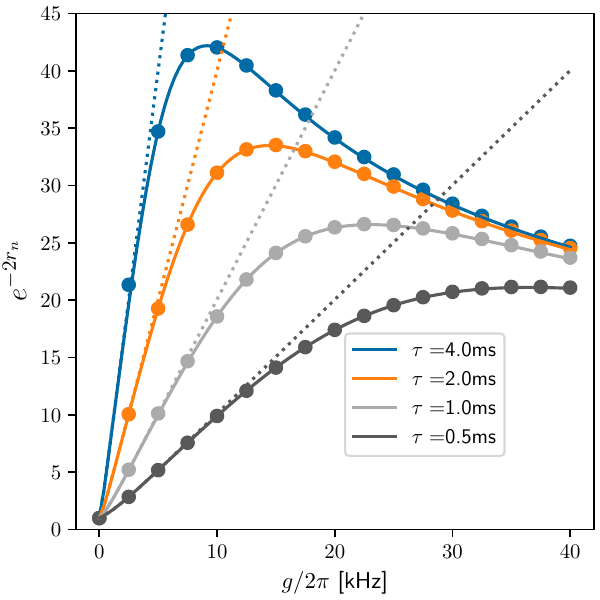}
    \caption{Amplification factor $e^{-2r_n}$ for the effective spin-spin interactions versus the parametric drive strength $g\equiv g_{nn}\sA_{nn}$ for various decoupling times $\tau$. Here, we set $\mu = 2\pi\times 3.045\,$MHz as in Fig.~4 of Ref.~\cite{ge2019PRL}. The solid (dotted) lines show the amplification when the detuning is not adjusted (is adjusted) to account for the mode frequency shift arising from the CR terms. Markers represent data from numerical simulations performed in Ref.~\cite{ge2019PRL}, which are shown in Fig.~4 of that paper.}
    \label{fig:counterrotating}
\end{figure}

Figure~\ref{fig:counterrotating} plots the amplification factor $e^{-2r_n}$ for the effective spin-spin interactions generated by the parametrically amplified spin-motion coupling, as the strength $g\equiv g_{nn}\sA_{nn}$ of the parametric drive is varied.  The colored curves correspond to different durations $\tau$ for which the SDF and parametric driving are simultaneously applied. The effective detuning $\delta_n'$ [Eq.~(\ref{eqn:detuning_eff_3d})] is chosen such that $\delta_n'=2\pi/\tau$, which ensures that the motion decouples at the end of the interaction time, leading to purely spin-spin interactions~\cite{ge2019PRL}. The solid lines plot Eq.~(\ref{eq:exp_rn_3d_cr}), where the detuning $\delta_n$ is not compensated to account for the CR terms and is hence obtained as $\delta_n' = \sqrt{\delta_n^2 - g^2}$. We note that the solid lines obtained from our analytical expression~(\ref{eq:exp_rn_3d_cr}) are in excellent agreement with the numerical results plotted in Fig.~4 of Ref.~\cite{ge2019PRL}, thereby validating our theory~\footnote{A perturbative estimate of the correction to $\delta_n'$ arising from the counter-rotating terms was derived in Ref.~\cite{ge2019PRL}. Our result coincides with that estimate for small values of the PA drive strength $g$. However, our result is applicable over a much wider range of $g$ values as evidenced by the agreement with the markers in Fig.~\ref{fig:counterrotating}.}. We plot these numerical results as markers in Fig.~\ref{fig:counterrotating} for a direct comparison. On the other hand, the dotted lines show Eq.~(\ref{eq:exp_rn_3d}), which assumes that $\delta_n$ is suitably shifted in the presence of the CR terms. The shifted $\delta_n$ can be obtained from $\delta_n'$ as $\delta_n' = \sqrt{(\delta_n + g^2/4\mu)^2 - g^2}$. The dotted lines in  Fig.~\ref{fig:counterrotating} demonstrate that the leading-order corrections from the CR terms do not present a fundamental limit to the amplification of spin-motion or spin-spin interactions provided the detuning is suitably adjusted.   

Finally, we note that our analysis only considers the leading order corrections from the CR terms and also does not exhaust other effects that can potentially limit the amplification. Importantly, the next-to-leading order correction from the CR terms would scale as $\sim \max(\delta_n, F_0, G_n)\cdot G_n^2/\mu^2$, whose effects we expect to be significant only at very large $g$ values beyond that considered in Fig.~\ref{fig:counterrotating}. Due to the reasonably small effect of the higher terms, we expect that other factors like the higher-order terms in the Lamb-Dicke expansion or mode frequency fluctuations might play a more important role in limiting the maximum achievable amplification.

\section{Conclusion and Outlook}
\label{sec:conclusion}
In this work, we have investigated the possibility of parametric amplification of spin-motion coupling in general 3D crystals. We find that unlike in lower dimensions, where the spin-motion coupling is amplified uniformly for each spin (i.e. `faithfully'), the amplification in 3D is in general unfaithful, and depends on the realization of the spin-motion Hamiltonian. However, with the specific choice of the phase-sensitive M\o lmer-S\o rensen configuration, the coupling to modes such as the center-of-mass mode can still be amplified faithfully. We also demonstrate a use-case for the non-uniform (i.e., 'unfaithful') amplification as well: In a bilayer crystal in a Penning trap, the light-shift configuration can be used to selectively amplify interactions within one layer and suppress them in the other, while maintaining the strength of the inter-layer coupling. Finally, we have re-analyzed the effects of counter-rotating terms on the limits of parametric amplification using Floquet theory. Our analysis indicates that the leading-order corrections do not hinder the amplification as severely as previously suggested, since they can be compensated for by simply changing the detuning of the SDF drive from the target mode.

Our findings have implications for trapped-ion systems in quantum sensing and simulation, where it is desirable to speed up entanglement generation without increasing the rate of decoherence. Our work demonstrates the feasibility of faithful PA in 3D crystals and also shows that unfaithful PA can be a resource to engineer non-trivial dynamics. Another setting where the latter can have interesting effects is when the longitudinal modes of an ion chain are used for spin-motion coupling~\cite{sackett2000Nature} instead of the transverse (radial) modes. In this case, assuming nearly uniform spacing between the ions, applying a parametric drive can lead to suppression of all ($2n$+$1$)th neighbour interactions (i.e. nearest, third-nearest etc.) and amplificaton of all $2n$-th neighbour (i.e second-nearest, fourth-nearest etc.) interactions. This tunability can allow for the generation of interesting spin phases. Finally, non-trivial mixing between nearby modes induced by the PA drive (e.g. modes 395, 396 in Fig.~\ref{fig:Anm_3d} (a)) could be used as a resource to generate unconventional spin-motion coupling Hamiltonians.

\section*{acknowledgments}
We thank Wenchao Ge and Allison Carter for feedback on the manuscript. We also thank Wenchao Ge for sharing the numerical simulation data that we have used in Fig.~\ref{fig:counterrotating} of this paper.
NN would like to thank Perimeter Institute and Boston University for support during this research. SH acknowledges partial support from the Austrian Science Fund (FWF) DOI 10.55776/F71 for the duration of this project. This work was supported by DOE  Quantum Systems Accelerator, ARO W911NF24-1-0128, and NSF JILA-PFC PHY-2317149. JJB and AMR acknowledge support through AFOSR grant FA9550-25-1-0080. AS acknowledges support through a New Faculty Initiation Grant (NFIG) from IIT Madras.

\appendix

\section{Form of Counter-Rotating Terms}
\label{app:CR_full}

In this Appendix, we derive the full Hamiltonian for the Light-Shift gate and the phase-sensitive M\o lmer-S\o rensen gate in the Lamb-Dicke approximation, including the counter-rotating terms.

\subsection{Light-Shift Gate}
\label{app:CR_full_LS}
The total Hamiltonian in the presence of parametric amplification is 
\begin{align}
\label{eq:A1}
    \hH &= \sum_n \omega_n \had_n \ha_n - \sum_{j} \frac{F_0}{\Delta k} \sin(\mu t - \Delta k(z_{0,j}+\hz_j))  \hs^z_j. \nonumber\\
    + &\Omega_p \cos(2\mu t-\theta)\sum_j \bigg((z_{0,j} + \hz_j)^2 - \frac{1}{2}(x_{0,j}+\hx_j)^2 \nonumber\\&- \frac{1}{2}(y_{0,j}+\hy_j)^2\bigg).
\end{align}

Under the Lamb-Dicke approximation, the SDF term can be expanded as follows, replacing $-\Delta k z_{0,j}$ with $\phi_j$:
\begin{multline}
    \hH_{\text{SDF}} = - \sum_{j} \frac{F_0}{\Delta k} \sin(\mu t +\phi_j)\hs^z_j \\+\sum_j F_0 \cos(\mu t + \phi_j)\hz_j \hSig^z_j. 
\end{multline}

Written out using the mode expansion (Eq.~(\ref{eqn:generalModeExpansion})):
\begin{align}
    \hH_{\text{SDF}} &= -\sum_j \frac{F_0}{\Delta k} \sin(\mu t +\phi_j)\hs^z_j \nonumber \\
    &+\sum_{j,n} F_0 l_n\cos(\mu t + \phi_j)(u^z_{nj}\ha_n+u^{z*}_{nj}\had_n)\hSig^z_j \nonumber.\\
    \hH_{\text{PA}} &= \Omega_p \cos(2\mu t-\theta)\bigg(\hO_{\text{PA,z}} - \frac{\hO_{\text{PA,x}}+\hO_{\text{PA,y}}}{2}\bigg),\\
    \text{where:}\nonumber\\ 
    \hO_{\text{PA,z}} &=\sum_j z_{0,j}^2  + 2\sum_{j,n}z_{0,j}l_n(u^z_{nj}\ha_n+u^{z*}_{nj}\had_n) \nonumber \\ &+\sum_{j,m,n}l_nl_m(u^z_{mj}\ha_m+u^{z*}_{mj}\had_m)(u^z_{nj}\ha_n+u^{z*}_{nj}\had_n).
\end{align}
$\hO_{\text{PA,x}}$ and $\hO_{\text{PA,y}}$ follow the same pattern as $z$. In the above expression, the $\hz^2$ terms in $\hO_{\text{PA,z}}$ give rise to mode couplings $\ha_n\ha_m$. Additionally, the $z_{0,j}^2$ term is operator-free, hence neglected from here on.\\ 

In going to a frame rotating with the SDF drive, via $\hU = e^{i\mu t\sum_n\had_n\ha_n}$, $\hH$ transforms as follows:
\begin{align}
     \omega_n\had_n\ha_n &\rightarrow (\omega_n-\mu)\had_n\ha_n \nonumber\\
     \ha_n &\rightarrow \ha_ne^{-i\mu t}.
\end{align}

We look at the different time-dependent terms in $\hH$. In $\hH_{\text{SDF}}$, both the drive and the mode operators rotate at $\mu$. This gives rise to stationary terms, terms that rotate at $\mu$, and terms that rotate at 2$\mu$:
\begin{align}
    \hH_{\text{SDF}} &= - \sum_{j} \frac{F_0}{\Delta k} \sin(\mu t +\phi_j)\hs^z_j\nonumber\\
    &+\sum_{j,n}\frac{F_0 l_n}{2}\bigg[u^z_{nj}\ha_n (e^{-i(2\mu t + \phi_j)} + e^{i\phi_j}) \nonumber\\
    &+ u^{z*}_{nj}\had_n (e^{i(2\mu t + \phi_j)} + e^{-i\phi_j})\bigg]\hSig^z_j.
\end{align}

Similarly, in $\hH_{\text{PA}}$ the PA drive rotates at $2\mu$ while the mode operators rotate at $\mu$. We write out the resulting expression for $\hH_{\text{PA,z}}=\Omega_p \cos(2\mu t-\theta)\hO_{\text{PA,z}}$:
\begin{align}
    &\hH_{\text{PA,z}} = \Omega_p\sum_{j,n}l_nz_{0,j}\bigg(u^z_{nj}\ha_n (e^{i(\mu t-\theta)}+e^{-i(3\mu t-\theta)}) + \HC\bigg)\nonumber\\
    &+\frac{\Omega_p}{2}\sum_{j,n,m}l_nl_m\bigg(u^z_{mj} \ha_m(u^z_{nj}\ha_ne^{-2i\mu t}+u^{z*}_{nj}\had_n) + \textit{h.c.}\bigg)\nonumber\\
    &\times\bigg(e^{i(2\mu t-\theta)}+e^{-i(2\mu t-\theta)}\bigg).
\end{align}
It can be seen that the $z_{0,j}\hz_j$ term has contributions of $\mu$ and $3\mu$. The $\hz^2_j$ term contains two mode expansions, and results in rotating contributions at $2\mu$ and $4\mu$, as well as stationary terms.\\

Finally, we can group together terms in the full Hamiltonian which have the same time-dependence, which in this case corresponds to rotations of frequency $n\mu$ (with $0\leq n\leq 4)$. We first look at the stationary terms with $n=0$, which are
\begin{align}
    &\hH_{\text{RWA}} = -\sum_n \delta_n \had_n \ha_n + \sum_{j,n}\frac{F_0 l_n}{2}(u^z_{nj}\ha_n e^{i\phi_j} + \textit{h.c.})\hSig^z_j \nonumber\\
    +&\sum_{j,n,m}\frac{\Omega_pl_nl_m}{2}\bigg[\ha_m\ha_n \bigg(u^z_{mj}u^z_{nj}-\frac{u^x_{mj}u^x_{nj}+u^y_{mj}u^y_{nj}}{2}\bigg) e^{-i\theta}\nonumber\\ 
    &+ \textit{h.c.}\bigg],
\end{align}
where the first line, containing terms from the free evolution and SDF, matches Eq.~(\ref{eqn:RWA_ODF}). Separating the $m=n$ and $m\neq n$ terms in the second line (PA terms), and defining expressions as in Eq.~(\ref{eqn:g_nm,A_nm defn}):
\begin{align}
    g_{nm} &= \Omega_p l_n l_m, \nonumber\\
    A_{nm} &= \sum_j \left(u_{nj}^z u_{mj}^z - \frac{u_{nj}^x u_{mj}^x + u_{nj}^y u_{mj}^y}{2} \right), \nonumber \\
    \implies \hH_{\text{PA}} &= \sum_{n} \frac{g_{nn}}{2}(\ha[2]_n e^{-i\theta}A_{nn} + \textit{h.c.}) \nonumber\\
    &+ \sum_{m\neq n}\frac{g_{nm}}{2}(\ha_n\ha_me^{-i\theta} A_{nm} + \textit{h.c.}).
\end{align}
Hence, Eq.~(\ref{eqn:RWA_PA}) is also obtained. Moving on to the counter-rotating terms, we get:
\begin{align}
\label{eq:CRderived}
    \hH_{\text{CR, SDF}} &= - \sum_{j} \frac{F_0}{\Delta k} \sin(\mu t +\phi_j)\hs^z_j\nonumber\\
    +&\sum_{j,n}\frac{F_0 l_n}{2}(u^z_{nj}\ha_ne^{-i(2\mu t + \phi_j)} + \textit{h.c.})\hSig^z_j.\nonumber\\
    \hH_{\text{CR, PA}} &= \sum_{j,n}\Omega_p l_n\bigg[\ha_n\bigg(z_{0,j}u^z_{nj}-\frac{x_{0,j}u^x_{nj}+y_{0,j}u^y_{nj}}{2}\bigg)\times\nonumber\\
    &(e^{i(\mu t-\theta)}+e^{-i(3\mu t-\theta)})+\textit{h.c.}\bigg]\nonumber\\
    &+\sum_{m,n}\frac{g_{nm}}{2}(B_{nm}\ha_n\ha[\dagger]_m+\textit{h.c.})(e^{i(2\mu t-\theta}+\textit{h.c.})\nonumber\\
    &+\sum_{m,n}\frac{g_{nm}}{2}(\ha_n\ha_me^{-i(4\mu t - \theta)}A_{nm}+\textit{h.c.}),
\end{align}
with $B_{nm} := \sum_j \left(u_{nj}^{z*} u_{mj}^z - \frac{u_{nj}^{x*} u_{mj}^x + u_{nj}^{y*} u_{mj}^y}{2} \right)$. 

\subsection{Phase-sensitive MS Gate}
\label{app:CR_full_PSMS}
For the phase-sensitive MS gate, we start similarly to Eq.~(\ref{eq:A1}). Here, the zeroth-order term coming from the Lamb-Dicke expansion of the phase-sensitive MS Hamiltonian (Eq.~(\ref{eqn:MS-Hamiltonian})) is:
\begin{align}
    \hH_{\text{motion-free}} =&\sum_j \frac{\Omega_{\rm eff}}{2}\left(\hSig_j^+ e^{i(\Delta k z_{0,j}-\mu t)} + h.c. \right)\nonumber\\
    + &\sum_{j} \frac{\Omega_{\rm eff}}{2}\left(\hSig_j^+ e^{i(\Delta k z_{0,j}+\mu t)} + h.c. \right),
\end{align}
where $e^{i\Delta k\hz_j}\approx 1$ to zeroth order. Simplifying, and replacing $-\Delta k z_{0,j}$ with $\phi_j$:
\begin{align}
    \hH_{\text{motion-free}} = \sum_j \Omega_{\rm eff}\cos(\mu t) \left(\cos(\phi_j)\hSig^x_j + \sin(\phi_j)\hSig^y_j\right).
\end{align}
Under the definition of local spin frames using the same Raman lasers (see Eq.~(\ref{eqn:ps_ms_ham}) and above), the spin operator $\cos(\phi_j)\hSig^x_j + \sin(\phi_j)\hSig^y_j$ is nothing but $\hSig^y_j$. Clearly, $\hH_{\text{motion-free}}$ contains only counter-rotating terms.\\

The remaining terms in the Hamiltonian are the same as in Eq.~(\ref{eq:A1}), with $\frac{F_0}{\Delta k}\to \Omega_{\text{eff}},\phi_j\to 0$ and $\hSig^z_j\to\hSig^x_j$ in the SDF. From here, the analysis proceeds exactly as before, giving us the following RWA Hamiltonian:
\begin{align}
    \hH_{\text{RWA}} &= -\sum_n \delta_n \had_n \ha_n + \sum_{j,n}\frac{\Omega_{\text{eff}} \Delta k l_n}{2}(u^z_{nj}\ha_n + \textit{h.c.})\hSig^x_j \nonumber\\
    &+\sum_{n,m}\frac{g_{nm}}{2}\left(\ha_m\ha_n A_{nm} e^{-i\theta}+ \HC\right).\label{eqn:H_psms_RWA_full}
\end{align}
Among the counter-rotating terms, only the form of $\hH_{\text{CR,SDF}}$ is different between the two gates:
\begin{align}
    \hH_{\text{CR, SDF}} &= \sum_j \Omega_{\rm eff}\cos(\mu t) \hSig^y_j \nonumber\\
    &+\sum_{j,n}\frac{\Omega_{\text{eff}} l_n}{2}(u^z_{nj}\ha_ne^{-2i\mu t} + \textit{h.c.})\hSig^x_j
\end{align}
Together with $\hH_{\text{CR, PA}}$ (see Eq.~(\ref{eq:CRderived})), this recovers Eqs.~(\ref{eqn:CR-terms-SDF}) and (\ref{eqn:CR-terms-PA}).

\end{document}